\documentclass[apjl]{emulateapj}
\usepackage{natbib}
\usepackage{color}
\usepackage{hyperref}
\newcommand{\so}{1H~0323+342}

\newcommand{\fe}{Fe~II}
\newcommand{\sy}{NLSy1s}
\newcommand{\nls}{NLSy1}
\newcommand{\lacc}{$\log{L_{bol}/L_{Edd}}$}
\newcommand{\acc}{L$_{bol}$/L$_{Edd}$}

\hypersetup{colorlinks=true, breaklinks=true, citecolor=blue, linkcolor=blue, menucolor=blue, urlcolor=blue}

\begin{document}

\title{The host-galaxy of the gamma-ray Narrow-line Seyfert 1  galaxy \so} 

\author{J. Le\'on-Tavares\altaffilmark{1, 2}, J. Kotilainen\altaffilmark{2}, V. Chavushyan\altaffilmark{1}, C. A\~norve\altaffilmark{3}, I. Puerari\altaffilmark{1}, I.~Cruz-Gonz\'alez\altaffilmark{4},  V.~Pati\~no-Alvarez\altaffilmark{1}, S. Ant\'on\altaffilmark{5, 6, 7}, A. Carrami\~nana\altaffilmark{1}, L.~Carrasco\altaffilmark{1},  J.~Guichard\altaffilmark{1}, K.~Karhunen\altaffilmark{8}, A.~Olgu\'{\i}n-Iglesias\altaffilmark{1}, J.~Sanghvi\altaffilmark{8}, J.~R.~Valdes\altaffilmark{1}}

\altaffiltext{1}{ Instituto Nacional de Astrof\'{\i}sica \'Optica y Electr\'onica (INAOE), Apartado Postal 51 y 216, 72000 Puebla, M\'exico     
  \email{leon.tavares@inaoep.mx}}
\altaffiltext{2}{ Finnish Centre for  Astronomy with ESO (FINCA), University of Turku, V\"ais\"al\"antie 20, FI-21500  Piikki\"o, Finland}
\altaffiltext{3}{ Facultad de Ciencias de la Tierra y del Espacio (FACITE) de la Universidad Aut\'onoma de Sinaloa, Blvd. de la Americas y Av. Universitarios S/N, Ciudad Universitaria, C.P. 80010, tCuliac\'an Sinaloa, M\'exico}
\altaffiltext{4}{ Instituto de Astronom\'{\i}a, Universidad Nacional Aut\'onoma de M\'exico, Ap. 70-264, 04510 DF, M\'exico} 
\altaffiltext{5}{ Instituto de Astrof\'{\i}sica de Andaluc\'{\i}a - CSIC, 18008 Granada, Spain}
\altaffiltext{6}{ CICGE, FCUP, Rua do Campo Alegre, P 4169-007 Porto, Portugal}
\altaffiltext{7}{ FCUL, Campo Grande, P 1749-016 Lisboa, Portugal}
\altaffiltext{8}{ Tuorla Observatory, Department  of Physics and Astronomy, University of Turku, 20100 Turku, Finland}

\begin{abstract}

We present optical and  near infrared (NIR) imaging data of the radio-loud~Narrow-line~Seyfert~1~galaxy  \so, which shows intense and variable gamma-ray activity discovered  by the \emph{Fermi} satellite  with the Large Area Telescope. NIR and optical images are used to investigate the structural properties  of the host galaxy of \so; this together with optical spectroscopy allowed us to examine its black hole mass.     Based on the  2D multiwavelength surface brightness modeling, we  find that, statistically,  the best model fit  is a combination of a nuclear component and a S\'ersic profile ($n \sim 2.8$).  However, the presence of a  disc component  (with a small bulge $n \sim 1.2$) remains also a possibility and  cannot be  ruled out with the present data. Although at first glance a  spiral-arm like structure is revealed in our images, a 2D Fourier analysis of the imagery suggests that such structure corresponds to an asymmetric  ring,  likely associated to a recent  violent dynamical interaction.   We discuss our results  on the context  of  relativistic jets production and galaxy evolution.

\end{abstract}

\section{Introduction}

The Large Area Telescope (LAT) onboard of the space mission  \emph{Fermi} has detected emission from a wide variety of astrophysical objects, from supernovae to active galactic nuclei (AGN). Of all the Fermi/LAT sources, more than 60\% are strongly beamed AGN \citep{2lac},  leading to the remark that  the most common type of sources in the $\gamma$-ray sky  are  AGN whose relativistic jet points towards the Earth (i.e.~blazars).  Two types of blazars are known, Flat-spectrum radio quasars  and BL Lacs. Despite  the  possible difference in their  nuclear environment that makes them look different  (i.e. the absence of  broad-emission lines), both blazar types share two common features: (i) the  presence of a prominent relativistic jet  responsible for the violent variability across the electromagnetic spectrum \citep[e.g.][]{leontavares_2011,leontavares_2012}, and  (ii) their host-galaxies have a  strong morphological similarity, they are giant elliptical galaxies \citep[e.g.][]{kotilainen_1998_bllacs,kotilainen_1998_fsrq,scarpa_2000,nilsson_2003,leontavares_2011_mbh}. The latter  has served as observational ground for the  paradigm that powerful relativistic jets can almost exclusively be launched from massive elliptical galaxies, which in turn links the presence of prominent relativistic jets to the  latest  stages of   galaxy evolution \cite[e.g.][]{toft_2014}.

 However,  the  detection of bright and variable $\gamma$-ray emission from narrow-line Seyfert type 1 galaxies (\sy)  by \emph{Fermi}  \citep{abdo_2009}, casts doubts on the exclusive relation between powerful jets and early-type galaxies.   Unlike $\gamma$-ray blazars,  \sy\ have always been found to be hosted by  late-type galaxies where the prevalence of pseudo-bulges has been suggested \citep{orban_2011,mathur_2012}.  The small black hole masses inferred from their  Balmer emission lines  (FWHM$_{H\beta}$ $< $2000 km s$^{-1}$), their voracious appetite indicated by the high accretion rates close to the Eddington limit, together with their  demography  detected so far in the local universe \citep[$z< 0.8$,][]{zhou_2006} and their compact radio structures \citep{Doi_2007,Doi_2011}, have served as arguments to  believe that \sy\ might be  AGN in an early phase of evolution \citep[see][for a review]{komossa_2008}.  The \sy\  detected  at $\gamma$-rays   appear to be  exclusively  \emph{radio-loud} \sy, a class that  conforms   only   7\% of the overall \sy\ population \citep{ komossa_2008}.  Although  blazar-like properties in radio-loud  \sy\ were previously reported \citep{yuan_2008},   being  $\gamma$-ray emitters was never anticipated due to their  small black hole masses and their host-galaxy type, namely,  spirals with high incidence of bars and ongoing star formation \citep{deo_2006, ohta_2007,sani_2010,caccianiga_2014}.

 Since the discovery of  $\gamma$-rays  from radio-loud \sy\ galaxies \citep{abdo_2009}, most of the attention has been paid to the study of their blazar-like emission \citep{abdo_2009_monitoring,liu_2010,foschini_2011,calderone_2011,dammando_2012,foschini_2012,jiang_2012,paliya_2013,paliya_2014},  while their relevance for the host galaxy jet paradigm  has been scarcely studied.  Studies  aimed to characterize the host-galaxies of  $\gamma$\sy\ are very scarce due to the lack of deep imagery.   So far, 1H~0323+342 ($z=0.061$; 1.177 kpc arcsec$^{-1}$ projected distance) is  the \emph{unique} $\gamma$\sy\ for which a host galaxy imaging  has been obtained \citep{zhou_2007, anton_2008}. As can be seen from Figure~\ref{fig:imaging}, at first glance the morphology of the host galaxy of \so\ is irregular and  shows  a low-surface brightness morphological peculiarity  that could be  associated to    a ring-like structure product of a recent merger event \citep{anton_2008} or rather to a one-armed  spiral structure \citep{zhou_2007}.

 Despite previous imaging studies of \so , no  structural modeling of its host galaxy has yet been conducted  to investigate whether (or not) the host-galaxy properties of $\gamma$ -ray \sy\ resemble those of their radio-quiet counterparts -- spirals  with pseudo-bulges.  Therefore, it is essential  to investigate  whether  or not  this new class of $\gamma$-ray sources  are likely to  be evolutionarily young objects (i.e.  small black hole masses),  as  has been claimed  for the overall population of \sy.  If this picture holds,  it  would imply that a young  AGN -- with its black hole is still growing under accretion rates approaching to the Eddington limit --   might   be able to launch (and collimate)  fully developed relativistic outflows at an early evolutionary stage. This  in turn would become an essential input to theoretical models for the formation of relativistic  jets, AGN feedback and galaxy evolution \citep[e.g.][]{dimatteo_2005, hopkins_2005, debuhr_2010}  

Motivated by the possibility that \so\  could be an outstanding case where a prominent relativistic jet  -- powerful enough to accelerate particles up to the highest energies-- uses a spiral galaxy as a launch pad,  we take advantage of  multi-filter   imagery   ($B$, $R$, $J$, and $K_s$)  to perform a systematic structural modeling of the \so\ host galaxy. We also undertook a contemporaneous spectroscopy  to  obtain a virial estimate of its black hole mass, thus allowing for a comparison between black hole masses obtained by different scaling relations in order to assess the suspected youth of the central black hole  of \so.

The observational data is presented in \S 2, along with the reduction processes  and  modeling. In \S 3 we present the structural analyses of the host galaxy  and these results are discussed in section \S 4 in the context of AGN activity  and black hole mass.   Our results are summarized in \S 5.  Throughout the manuscript we adopt  cosmological parameters  of $\Omega_{m}=0.3$, $\Omega_{\Lambda} = 0.7$ and a Hubble constant of   $H_{0} = 70$ Mpc$^{-1}$ km s$^{-1}$.

\section{Observations}

\subsection{Imaging}\label{sec:imaging}

The near-IR images of 1H 0323+342 were obtained in J and Ks bands at the 2.5m Nordic Optical Telescope (Roque de Los Muchachos, Spain) using the NOTCam. The NOTcam detector has a size of 1024 x 1024 pixels, and the pixel scale of the wide field camera used for the observations (0.234 arcsec pixel$^{-1}$) gives it a total field of view of  $\sim 4 \times 4$ arcmin$^{2}$. The observations were done on the nights of 23 and 25 of January 2013, with a seeing of ~1.0 arcsec on both occasions. A dithering pattern was used for the observations to allow accurate sky subtraction. The dithering step used was 40 arcsec, with the observation split into 50 second exposures. 1H 0323+342 was observed for 15 minutes in J and Ks bands on both nights, for a total exposure time of 30 minutes per filter.

\begin{figure*}
\includegraphics[width=\textwidth]{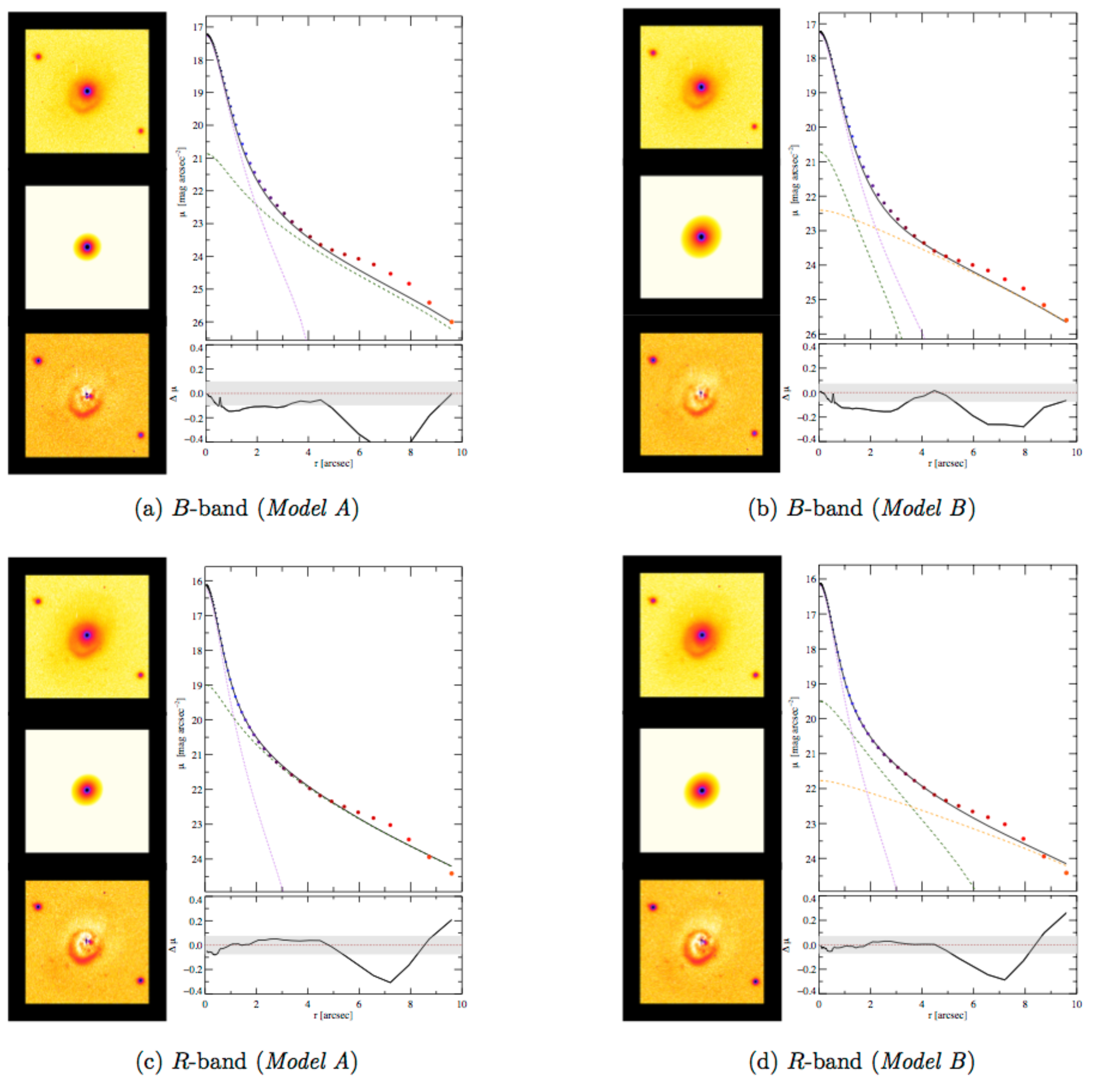}
 \caption{\small {2D surface brightness profile decomposition of \so\ at $B$- and $R$-bands for \emph{Model~A} (left) and \emph{Model~B} (right), respectively. The layout of each panel is as follows.  \emph{Top left sub-panel:}  the observed image at each band in a field of view of 20arcsec $\times$ 20 arcsec, North is left and East is up.  \emph{Middle left sub-panel:} Shows the model used to describe the surface brightness distribution, which turned out to be a combination of a nuclear unresolved component (PSF) and a S\'ersic model. \emph{Bottom left sub-panel:} The residual image. \emph{Top right sub-panel:} Radial profile of the surface brightness distribution of \so. The filled circles show the observations, the solid, pointed and  dashed lines represents the model, PSF and host galaxy, respectively. The exponential disc component is shown in orange color. \emph{Bottom right sub-panel:} Residuals. It should be noted that a significant residual is identified around 7 arc sec which coincides with the location of the ringed structure in the images of \so. }}
  \label{fig:imaging}
\end{figure*}

Data reductions were performed using the NOTCam quicklook package in IRAF\footnote{ IRAF is distributed by the National Optical Astronomy Observatories, which are operated by the Association of Universities for Research in Astronomy, Inc., under cooperative agreement with the National Science Foundation}. Bad pixels were masked using a mask file available in the NOTCam bad pixel mask archive. There was no need for dark subtraction to be performed, because, the utility of differential sky flats and sky-subtraction, will automatically subtract out the dark component in the image frames. For each day, two pairs of sky flats were observed for better estimation of normalized median combined master flat. Furthermore, they were interpolated over bad pixels, using bad pixel mask and corrected for the dc-gradient in differential images. For sky subtraction, a scaled sky template was produced from a list of dithered frames provided along with a master flat,  and then the sky was subtracted from it. Ultimately, using the field stars as reference points, the images were aligned to sub-pixel accuracy and combined to obtain a final reduced coadded image. Any cross-talk or horizontal strips were removed from these images without affecting the flux of the source. For zero point calibration, a cross-matching was performed between the photometry of field stars with the 2MASS database in the same filter. Fully reduced  NOT  $J$ and $K_s$images are displayed in panels c and d of Figure~\ref{fig:imaging}.

\subsection{Spectroscopy}\label{sec:spec}

Within the framework of a spectrophotometric  monitoring program of bright $\gamma$-ray sources \citep{patino_2013}, we undertook spectroscopic observations of \so\  using the  Boller \& Chivens long slit spectrograph on the 2.1m  Guillermo Haro Astrophysical Observatory  (GHAO) in Sonora, M\'exico.    The spectra were obtained  under  photometric weather conditions   (17 September 2012, 09 January 2013, 07 and 11 February 2013)  using a slit width of 2.5 arcsec. The spectral resolution was R=15 \AA$\,$ and R=7 \AA$\,$(FWHM) for the low resolution and the intermediate resolution spectra, respectively. The wavelength range for the three low resolution spectra is 3800-7100 \AA, while for one intermediate resolution spectrum the wavelength range is 4300-5900 \AA. The S/N ratio was $>$40 in the continuum near H$\beta$. To enable a wavelength calibration, HeAr lamp spectra were taken after each object exposure. Spectrophotometric standard stars were observed every night (at least two per night) to enable flux calibration.

The spectrophotometric data reduction was carried out with the IRAF package. The image reduction process included bias and flat-field corrections, cosmic ray removal, 2D wavelength calibration, sky spectrum subtraction, and flux calibration. The 1D spectra was subtracted taking an aperture of 6 arcsec around the peak of the spectrum profile. The spectra was also transformed to rest frame wavelength, applying a K-correction in the process as multiplying by a factor $(1+z)^3$, using the IRAF Task DOPCOR and the redshift obtained from measuring the peak of the H$\beta$ emission line. Top panel in Figure~\ref{fig:spec} displays a fully reduced optical spectrum of \so,  from which it can be  identified  the narrow H$\alpha$ and  H$\beta$ profiles  in addition to the Fe$_{II}$  bumps -- all features   that allow us to classify \so\ as a genuine NLSy1.

\begin{figure*}
\includegraphics[width=\textwidth]{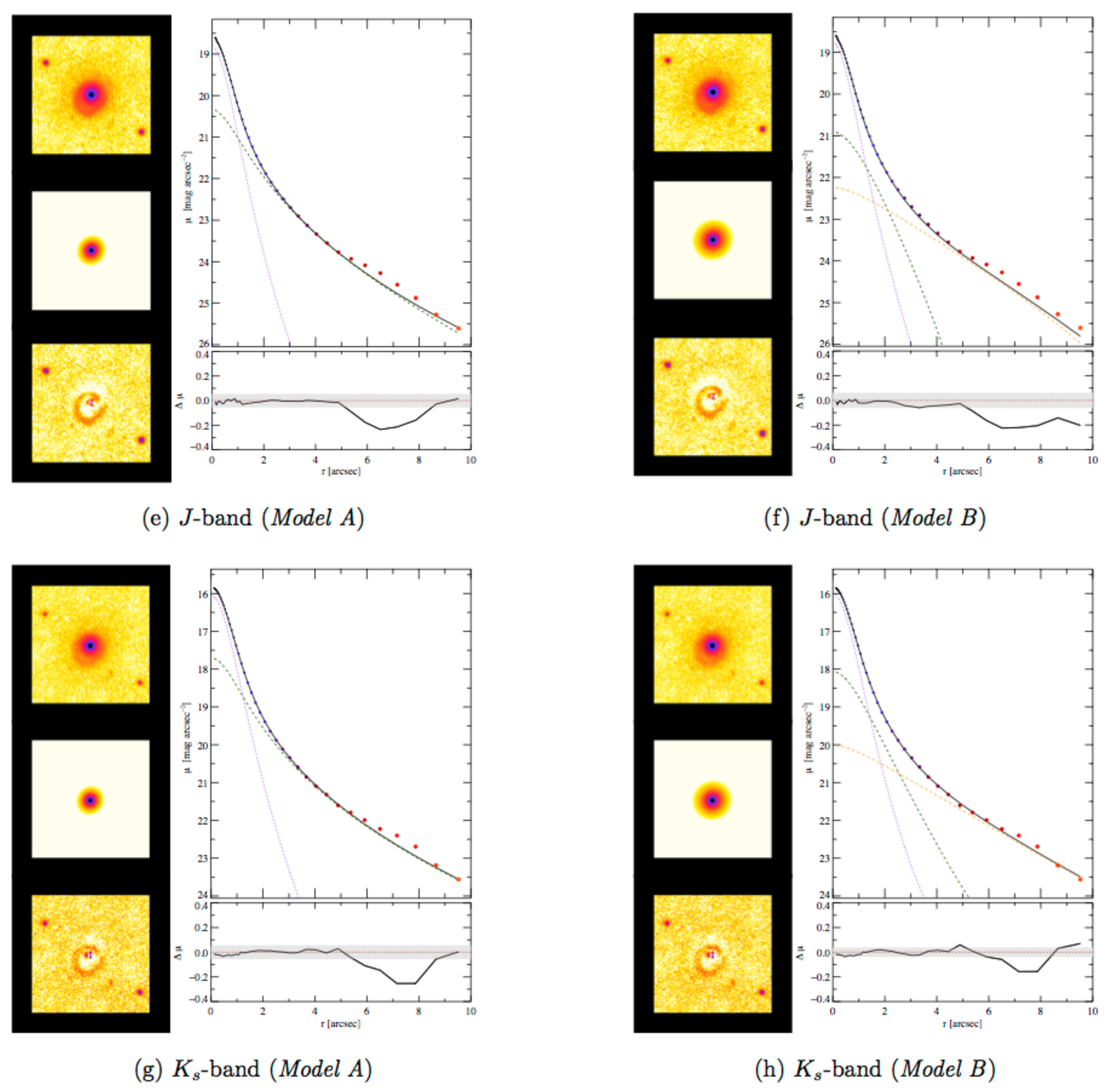}
\caption{The same as Figure~\ref{fig:imaging} but for $J$- and $K_s$-bands.}  
\end{figure*}

\begin{figure}
    \begin{center}$
    \begin{array}{cc}
    \includegraphics[width=\columnwidth]{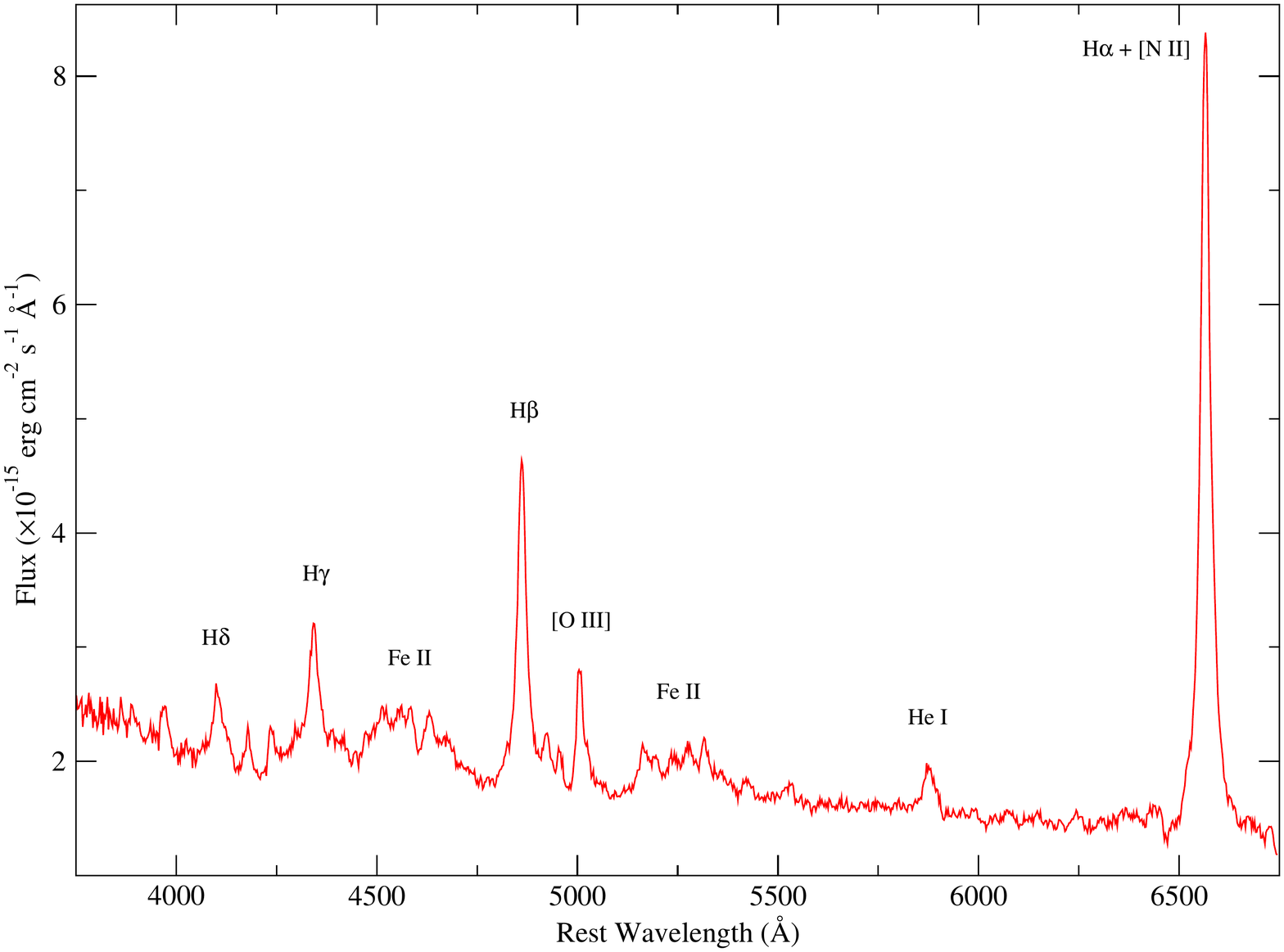}\\
    \includegraphics[width=\columnwidth,height=0.4\textwidth]{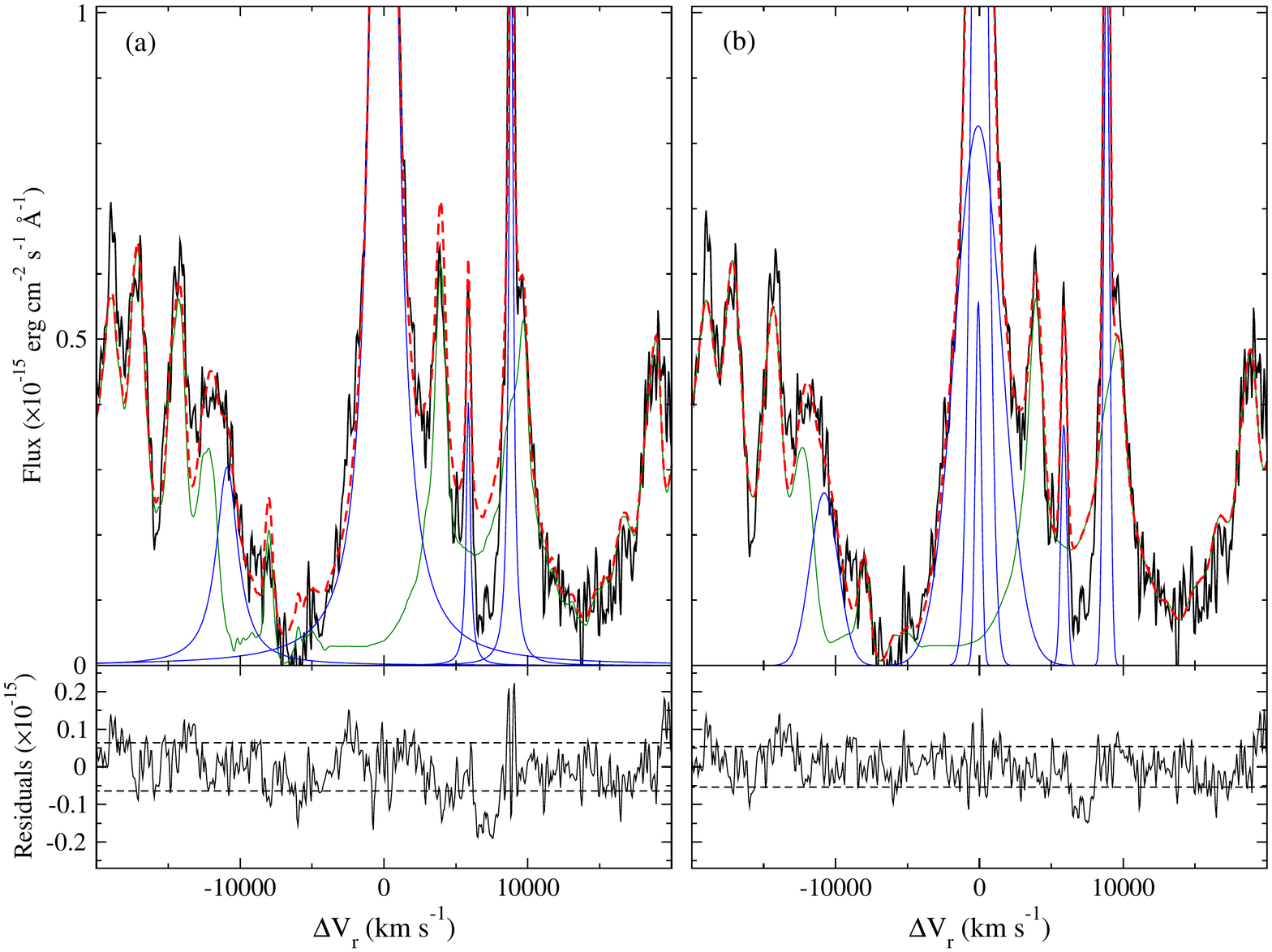}
    \end{array}$
    \end{center}
    \caption{\emph{Top:} Rest frame optical spectrum of \so, taken with the GHAO on 17th September 2012. Main spectral features are labeled.  \emph{Bottom:}  H$\beta$ profile: a) Single Lorentzian fitting, b) 3 Gaussian decomposition. In the bottom panels, the black line represent the original spectra; the green line is the fitted Fe II emission; the blue lines are the fitting to the emission lines (Lorentzians or Gaussians, respectively); and the red line is the total model. The bottom panels represent the residuals; the dashed lines represent the RMS values calculated in the range 4750-5100~\AA. The optical \so\  spectrum is available in a machine-readable form in the online journal.}
    \label{fig:spec}
    \end{figure}

\section{The Host galaxy}\label{sec:host}

\subsection{Photometric decomposition}

For the time being, 1H~0323+342  is  the \emph{unique} $\gamma$-ray  \sy\ for which a host galaxy imaging  has been obtained \citep{zhou_2007, anton_2008}. A peculiar structure  in the host galaxy  of \so\  has drawn attention,  and  is being  associated to either a spiral arm  or  a  ring structure.  In order to get a physical insight into  the host galaxy of \so\  and to determine its  evolution stage,  we use deep NIR imaging aiming to perform a  multi wavelength surface brightness host galaxy decomposition analysis.

In addition to our NIR imagery presented in Section~\ref{sec:imaging}, we  analyze  archival  $B$ and $R$ deep host galaxy images of \so\  taken also with  the NOT \citep{anton_2008}.  To study the structure of the  \so\ host galaxy  quantitatively, 
we use  the two dimensional  surface brightness model fitting code GALFIT  \citep{peng_2002}.  We considered two models  to fit the surface brightness of the host galaxy.  \emph{Model~A:}  single bulge component characterized with a S\'ersic profile; and  \emph{Model~B:}  bulge + disk component, where the disk is represented by an exponential disk. Both models include a   PSF to fit the  unresolved nuclear component associated to the AGN contribution (likely dominated by the non-thermal emission of the relativistic jets). The S\'ersic function \citep{sersic_1968} is described as
\begin{equation}\label{sersic}
I(R)= I_e \exp \left( -k \left[ \left( \frac{R}{R_e} \right)^{1/n} -1 \right] \right)~,
\end{equation}
where $I(R)$ is the surface brightness at the radius $R$, $I_e$ is the surface  brightness at effective radius $R_e$, which is defined as the radius where the galaxy contains  half of the light.  The S\'ersic index $n$  defines the  shape of the surface brightness function, while  $k$ is  a parameter coupled to $n$ in such way that $I_e$ is the surface brightness at the effective  radius. When $n=4$, the S\'ersic function becomes the de Vaucouleurs model \citep{devaucouleurs_1948}.

The exponential function \citep{freeman_1970} is defined as
 
\begin{equation}\label{exponential}
I(R)= I_0\exp\left(-\frac{R}{R_s}\right),
\end{equation}
where $I_0$ is the central surface brightness, and $R_s$ is the scale length of  the disk. The exponential function is a special case of the S\'ersic function when $n=1$. Since the images lack of high signal-to-noise-ratio (SNR) stars, we create a  point spread function (PSF) model for each filter by using  a Moffat profile \citep{moffat_1969,peng_2002}. To do this,  nearby stars were selected and fitted with GALFIT. Then, we created the PSF from  the fitted parameters.  The success of the PSF model  by using a Moffat profile is revealed  by the small amplitude, non-systematic variations in the residual image. In addition, Figure~\ref{fig:imaging} shows that our PSF fits properly in the central region of the galaxy.

To remove unwanted objects from the fitting, we use Sextractor \citep{bertin_1996} to create a mask image for the  fit. We took the ``segmentation'' image, and removed the mask area that  covers 1H 0323+342. We use the same  mask image for both models:  the S\'ersic  (\emph{Model~A}) and S\'ersic $+$ exponential  (\emph{Model~B}).   We need to provide GALFIT with initial parameters which were  were taken as initial guesses from a Sextractor run.   Magnitude,  effective radius, and axis ratio were taken from  the {\sf MAG\_BEST},  
{\sf FLUX\_RADIUS} and {\sc ELLIPTICITY} Sextractor parameters, respectively.  The value of PSF magnitude was chosen  slightly dimmer  than the S\'ersic magnitude. S\'ersic index was set initially  to $n=1.5$.  Then, all parameters in the S\'ersic  profiles and  the exponential disk were allowed to vary.  It should be noticed that setting (or fixing) the initial guess of the S«\'ersic index to $n = 4$  (typical value for blazar host galaxies)  or unmasking the spiral arm region, did not yield to better results. Table \ref{table:galfit} shows the best fit parameters for  the two galaxy models. Magnitudes listed in Table~\ref{table:galfit}  have been corrected for atmospheric and galactic extinction \citep{schlafly_2011}, and  included the corresponding K-correction based on \so\ host-galaxy color \citep{chilingarian_2010,chilingarian_2012}.

From Table \ref{table:galfit},  it can be gleaned that for both  models  --  at each filter -- the obtained $\chi^{2}_{\nu}$ values are  similar,  and  thus an evaluation  of  the fit goodness  merely based on these   values is not applicable to  assess  and compare models.  Moreover from an inspection of the smoothness of residuals,  it can be noticed that the residual peak-to-peak fluctuations percentage is similar for both models at each band.   Since both models are equally adequate in describing the profile of \so\  within all four bandpasses,  statistically speaking,   we are encouraged to  adopt the less complex model (fewer parameters): \emph{Model~A}. While one cannot completely exclude the possibility of a  disk component, after a visual inspection of the residual image and surface brightness profiles (see Figure~\ref{fig:imaging}),  there is no obvious presence of a   ``missing component".  Despite the single spiral arm structure left in the residual image, the residuals are dominated by  randomness.

\subsection{2D Fourier analysis}

In order the get further insight into the structure we observe in the images of \so, we performed a 2D Fourier analysis based on logarithmic spirals. This technique was originally proposed by \citet{kalnajs_1975} and was largely used for morphological analysis \citep[][among others]{considere_1982, considere_1988, puerari_1992,block_1999,puerari_2000}. Basically, the Fourier amplitude of a deprojected image of a given disk galaxy is calculated by

$$A(p,m)={1\over D}\int_{u_{min}}^{u_{max}}\int_{-\pi}^{+\pi} I(u,\theta)exp[-i(m\theta+pu)]dud\theta$$

\noindent where $D=\int_{u_{min}}^{u_{max}}\int_{-\pi}^{+\pi} I(u,\theta)dud\theta$, $I(u,\theta)$ is the intensity of the image at ($u$, $\theta$), $u=\ln(R)$, $R$ and $\theta$ are the polar coordinates, and $u_{min}$ and $u_{max}$ are the radial limits in the $\ln(R)$ direction. $p$ and $m$ are the frequencies related to $\ln(R)$ and $\theta$, respectively. The azimuthal frequency $m$ represents the number of arms, and $p$ is related to the pitch angle $P$ by $\tan(P)=-m/p$. We must note that different signs of $p$ represent spiral structures with opposite windings. e.g., for $m=2$ structures, $p<0$ represents spirals with a ``Z'' shape, while $p>0$ represents ``S'' shaped spirals.

\begin{figure}[!ht]
  \begin{center}\includegraphics[width=\columnwidth]{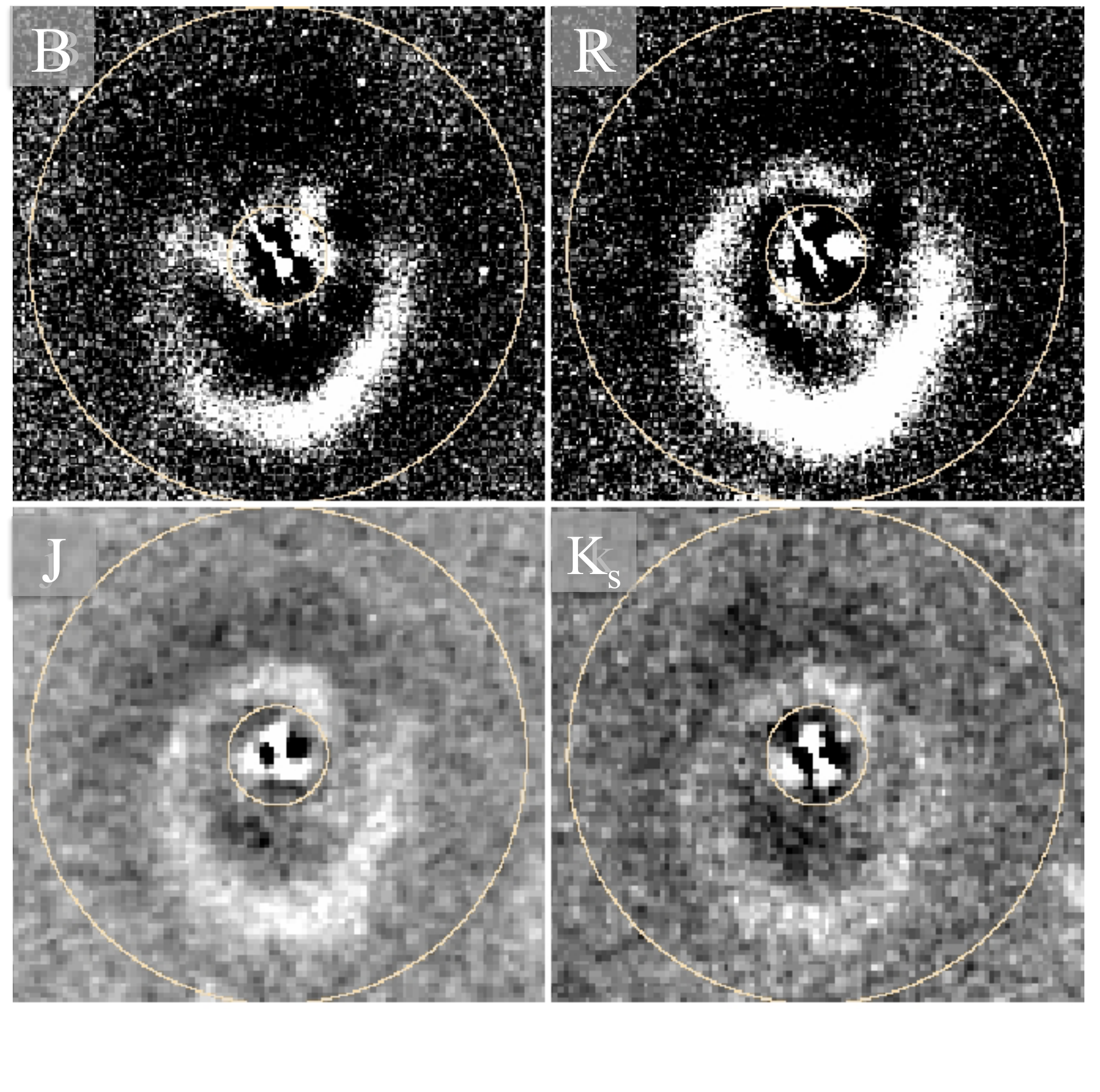}\end{center}
  \caption{Mosaic of  $B$, $R$, $J$, and $K_s$  deprojected images of \so. The circles are drawn at $R_{min}$ and $R_{max}$ and enclosed the annular region which is analyzed with the 2D Fourier technique.}
  \label{mosaic_deprog}
\end{figure}

One can easily note that in a $\ln R  \times \theta$ plane, a logarithmic spiral will be represented as inclined straight lines (positive or negative inclination depending on the sign of $p$). For a single $m=1$ armed structure, the 2D Fourier power spectra will show a {\sl single} peak for this component (see for example, the spectra for M~31 in \citet{kalnajs_1975,considere_1982, puerari_1992}. As we will see later, \so\ has an $m=1$ spectrum showing 2 peaks, one at $p<0$ and other at $p>0$, more representative of an elliptical structure, probably a ring.

\begin{figure}[ht]
  \begin{center}\includegraphics[width=\columnwidth]{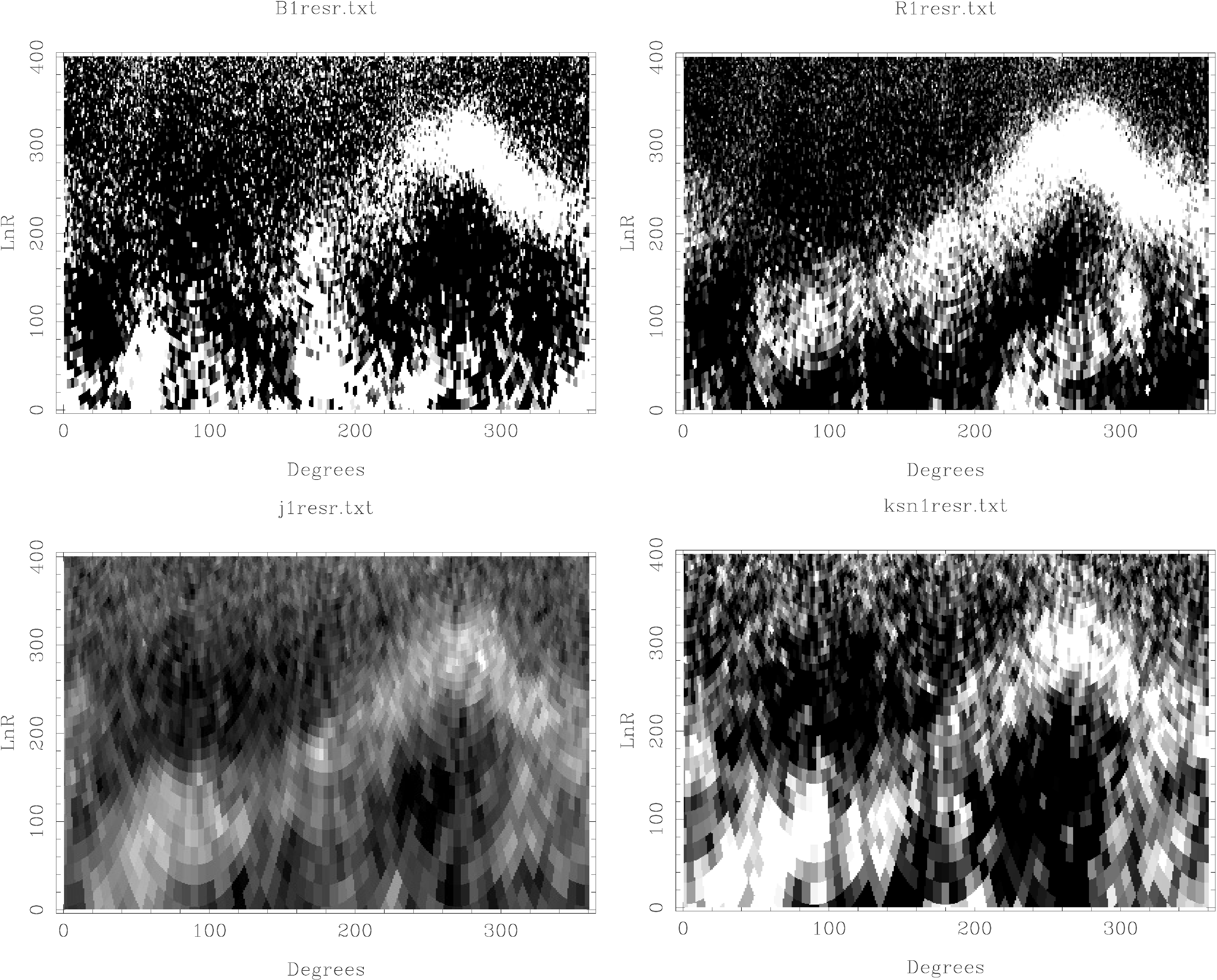}\end{center}
  \caption{Mosaic of the images, now in the $\ln(R)$ vs. $\theta$ plane. The horizontal axis is $\theta$ from 0 to $2\pi$ and the vertical one is $\ln(R)$ which range is $\ln(R_{min})$ to $\ln(R_{max})$. Notice the saw-tooth shape, more representative of a ringed structure than standard spiral arms. Notice also the bright structure at the $R$ image, at inner radii for $220^{\circ}<\theta<320^{\circ}$.}
  \label{mosaic_lnr}
\end{figure}

We developed a 2D Fourier software and applied it to the deprojected $B$, $R$, $J$ and $K_s$ images of \so. Before doing the Fourier calculation, we have subtracted the axisymmetric component. This does not change any power in $m's$ different than $m=0$. In Figure \ref{mosaic_deprog} we show a mosaic of the four filter  images. We can easily notice a curved structure which appears as a broken ring or an $m=1$ spiral arm. The $R$-image shows a strong structure just below  and closer to the center, which is hardly visible in the other filters. The circles in these images represent the radii $R_{min}$ and $R_{max}$. In Figure \ref{mosaic_lnr} we present a similar mosaic, now in the ($\ln R,\theta$) plane. The saw-tooth shape we can notice in these plots are more representative of a ringed structure.

Recently, \citet{savchenko_2013} have measured the spiral arms pitch angle variation (in radius) for a sample of 50 non-barred (or weakly barred) grand-design spirals. They found that most of the spiral arms cannot be described by a single value of the pitch angle. Instead, the galaxies show decreasing pitch angle values for increasing radii (more tightly wound spirals in the external part of the disk). For \so\ the situation is quite different. The two structures seen in each panel of Figure~\ref{mosaic_deprog} are represented by spirals with pitch angles that differs in sign, i.e., one structure has pitch angle around $P=+11^{\circ}$, while the other one shows $P=-18^{\circ}$. In the \citet{savchenko_2013} study, 2/3 of their sample shows pitch angle variations exceeding 20\%, but the arms they analyzed never changed the sign of the pitch angle value.

It should be noticed that a number of disk galaxies present resonant rings or pseudo rings -- see \citet{buta_2011} for an intensive discussion about ring and pseudo ring morphology. These rings are generally placed at some resonant radius, they are nearly circular, with the galaxy nucleus placed at the ring center, and sometimes, a bar. Therefore, the observed lopsidedness of  the structure  in \so\ is difficult to explain with a resonant ring.

Two scenarios of interactions can be suggested.  We could be witnessing the process of a small companion spiraling to the center of a massive galaxy. The particular geometry and the actual time of the orbit could be showing us that lopsided structure. From the high resolution (residual) image  presented in  \citet[][their Figure 2]{zhou_2007}, 
the structure does look like a spiral as it connects to and appears to originate from the nucleus. The structure shows an asymmetric brightness distribution, being broken in some azimuthal position. The abrupt decrease of the pitch angle is also a striking characteristic. Furthermore, some minor mergers could trigger the activity in galaxies \citep[e.g.][]{hernquist_1995}.

Other alternative is that the structure results from a passage/collision as those forming ring galaxies like  the Cartwheel \citep{struck_1993, athanassoula_1997,curir_2006,fiacconi_2012}.  The Cartwheel is shown to be an example of central, almost perpendicular passage of a spherical companion through a disk. Other encounters, with a companion passing through the disk in a off-center and/or a non perpendicular orbit, result in rings which are not so symmetric as Cartwheel. Indeed, very high simulations by \citet{fiacconi_2012} (see their Figure 2) show a variety of morphologies resulting from different interaction geometry.  However, we note that the size of the bulge derived from \textit{Model A} is  larger than the ring, so the ring has to be enshrouded by the bulge. Such a geometric configuration, if real, would present an interesting case as it is rarely seen.

\begin{figure}[ht]
  \begin{center}\includegraphics[width=\columnwidth]{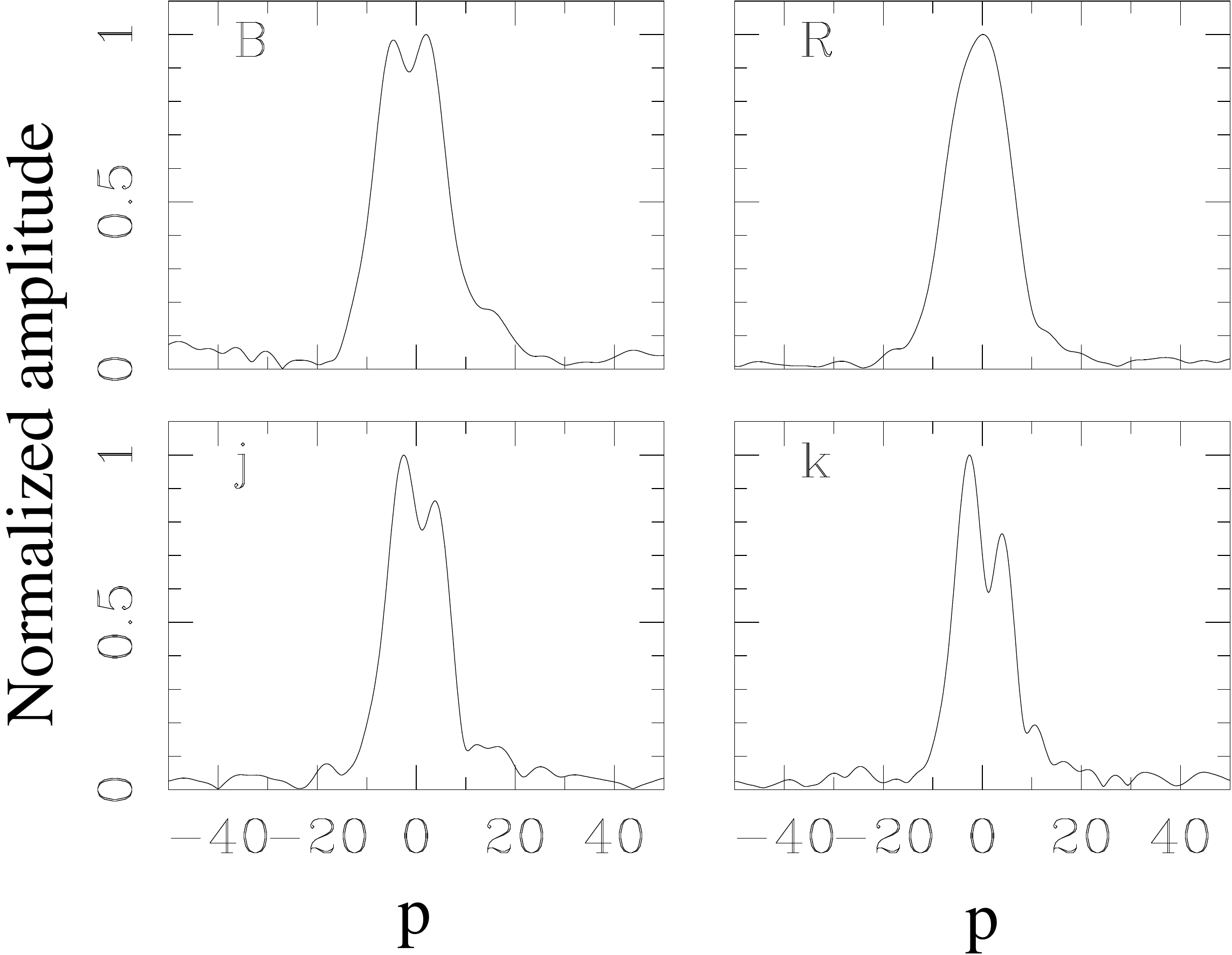}\end{center}
  \caption{Spectra for $m=1$. The two peaks, one for $p<0$ and the other for $p>0$ are representative of a ring structure. An inner lopsided structure seen in the $R$ filter affects the spectrum and provides high amplitudes for $p$ around 0, which erases the ``valley'' seen in the other filters.}
  \label{figure_spectra}
\end{figure}

 Finally, in Figure \ref{figure_spectra} we show the $m=1$ Fourier spectrum we have calculated. As expected, they represent a ringed structure. We can note the $p<0$ peak, representing a ``('' shape, and the $p>0$ peak, representing a ``)''structure. For the $R$ filter, the strong structure we see just below the center probably affects the spectrum for $m=1$ at $p=0$, erasing the ``valley'' we see between the peaks in the other filters.

The ringed structure detected via 2D Fourier transform on the basis of logarithmic spirals does not necessarily  mean that the real structure is indeed a ring on the disc plane. As discussed above, for a given geometry, some interaction in which the companion is disrupted when orbiting the disk can generate debris which could be detected as a projected ringed structure. The essential ingredient to solve the problem will be to get kinematical data from the ringed structure to decide if it is indeed an off centered ring or a disrupted intruder.

 \section{Discussion}

\subsection{The black hole mass }\label{sec:bh}

In this section we examine the black hole mass  in order to investigate whether (or not) \so\   is an evolutionarily young object as suspected for the overall population of  \sy.  Since  a black hole accelerates nearby matter,   its  mass can be estimated by  modeling the kinematics of either  nearby gas \citep[within the BLR; e.g. ][]{kaspi_2000} or stars in its vicinity \citep[inside the bulge of its host galaxy; e.g.][]{magorrian_1998,tremaine_2002}. However,   recent results have shown that in radio-loud AGN, the gas of the BLR can be accelerated and ionized by the non-thermal emission from the jet \citep{leontavares_2010,leontavares_2013,arshakian_2010}, which in turn might introduce uncertainties in the black hole mass estimated by  assuming that  the BLR clouds follow a virialized motion.  Bearing this in mind, we consider two indirect methods to estimate the mass of  the black hole of \so: (i) via the width of its broad emission lines,   and (ii) via the luminosity of its bulge.

Despite several studies on the emission line properties of \sy, there is no consensus yet on how  the profiles of the emission lines in \sy\ should be properly characterized.  Because of the  cusped peaks and broad wings of  H$\alpha$ and H$\beta$ emission lines, some works  have suggested that \sy\  emission lines have  Lorentzian profiles \citep[e.g.][]{veron_1980,veron_1981a,veron_1981b,moran_1996,Goncalves_1999,veron_2001}. Nevertheless, more recent studies \citep[e.g.][]{rodriguez_ardila_2000,dietrich_2005} found no evidence of Lorentzian profiles  and instead suggest that the profiles of \sy\ can be well represented by using two (or three) Gaussian components. 
Building on the above studies, we  follow  two different approaches to characterize the H$\beta$ emission line in \so: (a) single Lorentzian fitting,   and  (b) three Gaussian decomposition.

The fitting is performed using the task SPECFIT  \citep{specfit} from the IRAF package. The steps we followed for identification, de-blending, and line measurements  in each spectra are:   {\it The continuum} -- We adopted a single power-law fit to describe a local continuum using the continuum windows around 4200 and 5400 \AA. {\it Fe II emission} -- We adopt the template of the NLSy1 prototype  I~Zw~1 \citep{marziani_2003}  to characterize the Fe II emission in the optical band. {\it H$\beta$ emission line} -- {\it Case (a)} -- The entire H$\beta$ profile was fitted with a single Lorentzian, as well as the He II $\lambda$4686 emission line; the [O III] $\lambda\lambda$4959,5007 emission lines have the same FWHM and are fitted with a single Gaussian each. {\it Case (b)} -- We assume that a single value of FWHM is adequate to fit the narrow components of the H$\beta$ and [O III] lines ([O III] $\lambda$5007 FWHM was adopted); while the broad component of H$\beta$ was fitted using two independent Gaussians; the He II $\lambda$4686 emission line was fitted with a single Gaussian.  {\it Final Fitting}-- A simultaneous fitting of all the components is performed by Chi-square minimization using a Marquardt algorithm. Example of the fits obtained for the intermediate resolution spectrum (2013 Feb 11) using these two approaches can be seen in the bottom panels of Figure~\ref{fig:spec}.

We make use of the relations presented in \citet{vestergaard_2006} to estimate the  black hole mass of \so\ via the  continuum (5100~\AA) and H$\beta$ luminosities. We consider the later  scaling relation because it is very likely that, during an outburst episode,  the  optical continuum emission of \so\  could be dominated by the non-thermal radiation from its relativistic jet --  responsible for the production of gamma-ray photons.  The relations to estimate black hole mass  using the H$\beta$ and 5100~\AA\ continuum luminosities are expressed below,

\begin{equation}
\scriptsize{
\log M_{BH}(H\beta) = \log \bigg[\bigg(\frac{FWHM(H\beta)}{1000\,km\,s^{-1}}\bigg)^2 \bigg(\frac{\lambda L_{\lambda}(5100\,\AA)}{10^{44}\;erg\,s^{-1}} \bigg)^{0.5}\bigg] + 6.91}
\label{conti}
\end{equation}

\vspace{5pt}

\begin{equation}
\scriptsize{
\log M_{BH}(H\beta) = \log \bigg[\bigg(\frac{FWHM(H\beta)}{1000\,km\,s^{-1}}\bigg)^2 \bigg(\frac{L(H\beta)}{10^{42}\;erg\,s^{-1}} \bigg)^{0.63}\bigg] + 6.67}
\label{linea}
\end{equation}
and the corresponding black hole mass estimates are shown  in  columns (5) and (6) of  Table~\ref{table:bh_spec}, respectively.

We also make use of the relation between the black hole mass and the bulge luminosity (M$_{BH}$-M$_{bulge}$)  to  compare and  test the consistency of  our black hole mass estimates. Since we have performed the photometric decomposition of the  host galaxy of \so\ in four different filters,  we can make use of at least two M$_{BH}$-M$_{bulge}$ relations calibrated for two different filters, namely $R$ and  and $K_S$   \citep{graham_2007}. The expression we used to estimate the black hole mass of \so\ via the magnitude of its bulge  are listed next,

 \begin{equation}\label{eq:mR}
 log~M_{BH} = -0.38(M_{R} + 21) + 8.12
 \end{equation}
 
\begin{equation}\label{eq:mK}
log~ M_{BH} = -0.37(M_{K_s} + 24) + 8.29
 \end{equation}
and the associated intrinsic scatters are about 0.3 dex.  The black hole mass estimates for \so\  using the above expressions  and the bulge magnitudes from \textit{Model~A}  (listed in Table~\ref{table:galfit}) are    {\it log~M}$_{BH} (R) =8.5$ and   {\it log~M}$_{BH} (K_{s}) =8.7$;  hereafter we shall considered a mean value of {\it log~M}$_{BH} (M_{bulge}) =8.6$ . The later values are consistent with black hole mass estimates obtained by  other derived expressions in the NIR \citep{vika_2012}.  On the other hand, if we consider the bulge properties derived for \textit{Model B}, we obtain  {\it log~M}$_{BH} (R) =8.1$ and   {\it log~M}$_{BH} (K_{s}) =8.3$, which yields a mean value of 8.2. \textit{Model~B} yields a smaller mass than \textit{Model~A}, as expected for a smaller bulge.

 As can be seen from Table~\ref{table:bh_spec},  the black hole mass  estimates of \so\ obtained  by using the single-epoch spectra scaling relations  \citep{vestergaard_2006}   are significantly lower  (by about one order of magnitude)  than those estimated via the bulge magnitude (found in both \textit{Model A} and \textit{Model B}) . This apparent disagreement has been previously noticed in  several   studies on \sy\ \citep[e.g.][]{mathur_2001,grupe_2004,komossa_2007,ryan_2007} and has been taken  as an observational evidence for  the intrinsic youth of \sy. It has been suggested that  \sy\  do not follow the M$_{BH}$ - $\sigma$ relation \citep[e.g.][]{mathur_2001},  probably because they are still accreting material and therefore their bulges have not had enough time to virialize after a possible galaxy encounter.   It should be noticed that independent evidence for small black hole masses in NLSy1 has been suggested from X-ray variability analysis \citep{nikolajuk_2009,ai_2011,ponti_2012}, albeit only for radio-quiet sources.

 An alternative explanation for such a  disparity in black hole mass estimates (and the narrow H$\beta$ profiles in \sy) assumes the presence of a flat (i.e. disc-like) broad-line region \citep{osterbrock_1985,collin_2004,decarli_2008,decarli_2011}. In this scenario,  the H$\beta$ profile depends sensitively on the  inclination of the BLR. If the viewing angle between the line of sight and the  jet  axis is large (e.g.  $\theta \sim 90^{\circ}$; an edge-on disk), then  emission lines arising from a flat BLR will appear broad, even double-peaked \citep[e.g.][]{shapovalova_2010_ngc4151,shapovalova_2010}. The red and blue wings  in the profile are associated to the receding and approaching regions of the rotating disk, respectively. As the viewing angle reduces,  the two wings of the emission line  get closer together until they merge into a  very narrow profile for a  disk  seen face-on ($\theta = 0^{\circ}$).  Following the  methodology presented in \citet{decarli_2008} and assuming a   viewing angle  typical of gamma-ray emitters ($\theta \leq  5^{\circ}$), we obtain an estimate for the black hole mass of    {\it log~M}$_{BH} (R)  \ge 8 $ . This  value is in fair agreement with the black hole mass estimated via the bulge luminosity and  is also consistent with  black hole masses estimated  in a sample  of radio-loud  \sy\ via modeling of the optical and UV data with an accretion disc  model  \citep{calderone_2013}. 
 
\subsection{The active nucleus}\label{sec:agn}

What  type of AGN is  \so? The measurements of its nuclear optical spectra features --  $FWHM_{H\beta} \sim 1600$ km s$^{-1}$, R$_{4570} = Fe II / H\beta \sim 2$,  R$_{5007} = [ O III ]/ H\beta \sim 0.12$ , allow us to  classify it as a   NLSy1 \footnote{R$_{4570}$ and  R$_{5007}$ measurements have been taken from \citet{zhou_2007}. We will explore  variability of these spectral features in a forthcoming article. }.  At the same time, its  non-thermal emission signatures -- presence of a  core-jet structure \citep{anton_2008,wajima_2014} and  detection of intense and variable high energy photons in the \emph{Fermi} energy bands \citep{abdo_2009}, strongly  suggest that  a blazar nucleus might be at work.     Previous works  found an anti correlation between radio-loudness and Eddington ratio  (\acc) \footnote{The Eddington ratio is defined as the ratio between the bolometric  and Eddington luminosities, and is often used as a dimensionless proxy for the accretion rate.} in a sample of Seyfert 1 and NLSy1 galaxies,   \citep{ho_2002,greene_2006}.  The  trend of  increasing  radio-loudness with decreasing Eddington ratio suggests  that  \nls\ sources --  accreting close or above the Eddington~limit (i.e. \lacc\ =1), should be  radio-quiet ($\log{R} <0$). Interestingly, \so\ does not follow this trend at all, with an extreme  radio-loudness  \citep[$\log{R} = 2.4$;][]{doi_2012}, and its Eddington ratio  \lacc\ $ > -1.9$ \footnote{ We estimate bolometric and Eddington luminosities  as  L$_{bol} \sim 9  \lambda L (5100\AA)$  and L$_{Edd}~=~1.3 \times 10^{38}  ( M_{BH} ) $, respectively. The mean estimated black hole masses  (see section~\ref{sec:bh}) are  $\log{M_{BH,~virial} }= 7.1~M_{\odot}$ and  $\log{M_{BH,~bulge}} = 8.6~M_{\odot}$.  Eddington ratios  for \so\ range from  $\log($L$_{bol}$/L$_{Edd(bulge)})\sim-1.9$ to  $\log($L$_{bol}$/L$_{Edd (virial)})\sim -0.4$. } is significantly higher  to the value predicted by the above mentioned relation (\lacc~=~-4.3). \citet{yuan_2008} already noticed that radio-loud NLSy1 do not follow the inverse relation between \acc\ and radio-loudness.  This, might suggest that radio-loud \nls\ depart from the population of Seyfert and radio-quiet \nls\ galaxies. Although there is evidence of radio-loud sources having high Eddington rates \citep{sikora_2007}, those are classified as quasars and might be intrinsically different from Seyferts.  Therefore, any comparison between \so\  (an extreme radio-loud \nls) in the context of  the Seyfert and radio-quiet \nls\  galaxy population  should be handled with caution.

If we were to favor the disk-like BLR and the pole-on geometry in  \so,  then we would be implicitly  assuming  the bulge magnitude as  the most reliable proxy for the black hole mass ($\log{M_{BH,~bulge}} \sim 8.6~M_{\odot}$, consequently   $\log$~L$_{bol}$/L$_{Edd(bulge)}\sim-1.9$).  In this scenario,  \so\ would represent a gamma-ray emitter which  black hole mass,  luminosity and \acc\  are average among extragalactic high-energy sources.  Thus, there is no need of invoking \acc\ close (or above) to the Eddington limit.   However,  at this point the discerning reader might ask:  Could \so, along with the afore mentioned \acc\ and black hole mass within the blazar range, produce  the  intense \fe\ emission observed?

Before trying to address the above question, we should   bear in mind the following issues: (i) Although \fe\ emission is prominent  and a casting signature in \sy\ \citep{veron_2001},  this feature it is not exclusive to \sy\ and can be  found in radio-loud AGN as well \citep[e.g.][]{jackson1_1991, miller_1993, torrealba_2012}, albeit with lower intensity; (ii) Since the \fe\ emission is complex, with a large number of lines blending with each other and with adjacent emission lines,  the main physical driver(s) behind the  production of such lines remains under debate.  A principal- component analysis  of quasar spectra suggested  a tight connection between  \acc\ and  \fe\ emission \citep{boroson_1992,marziani_2001}. However, details on this connection are not yet fully understood. The \fe\ emission is believed to be a major coolant of the BLR \citep[e.g.][]{garcia-rissmann_2012},  and different gas heating mechanisms have been suggested in the past.  It has been proposed that  \fe\ emission could be generated via photoionization mechanisms and  arise from a flatter distribution (i.e. a disk) \citep[e.g.][]{marziani_2001}.

 Alternatively, the model by \citet{joly_1991},  based on the scheme presented in \citet{Norman_1984},   proposes that \fe\ emission could be produced when the jet interacts with  its environment.   Nevertheless, both models assume  that \fe\ emission is anisotropic and thus orientation dependent. In the case where the \fe\  originates in a disk, the \fe\ emission would be increased when such disk would be seen face-on (or the galaxy seen pole-on). On the other hand, in the jet model, emission could be boosted due to the orientation angle of the jet.  Then, the excess of \fe\ seen in the optical spectra of \so\  could be a consequence of the anisotropic emission.  \citet{leontavares_2013} found that the \fe\  in the optical spectra of the  bright gamma-ray source 3C~454.3, responds to the variation of the non-thermal optical continuum emission. This in turn could be considered suggestive  that in sources with prominent relativistic jets, as in \so, the degree of anisotropy in the \fe\ emission might be significant,  and  that the ionizing non-thermal continuum  emission from the jet  can play a major role in the production of \fe\ emission.

\subsection{Host-galaxy and environment}

In the current \sy\ paradigm, the relative youth associated to these sources is accountable for the following observational properties: (\emph{i}) late-type host galaxy, (\emph{ii}) under massive black holes  and  (\emph{iii}) high accretion rates approaching to the Eddington limit.  However, as  previously discussed (see sections~\ref{sec:bh}, \ref{sec:agn}),  properties \emph{ii} and \emph{iii} are prone to uncertainties in the methods used to estimate the mass of the central black hole. Therefore,  the characterization of the host-galaxy arises as the most unbiased method to verify  the intrinsic youth of \sy.

 However, given the limited  spatial resolution and depth of our imagery, we cannot rule out completely the presence of  a disc component in the host-galaxy of \so. Moreover, the disc structural parameters derived for \emph{model~B} are within the range of disc parameters  reported  in a sample of radio-quiet NLSy1 galaxies found to be hosted by pseudo-bulges \citep{mathur_2012}.   The presence of pseudo-bulges in \sy\ has been  previously proposed  \citep[e.g.][]{orban_2011, mathur_2012}, in this context, the morphological parameters (i.e. $ n < 2 $) derived for  \emph{model~B} (see Table~\ref{table:galfit})  might suggest the presence of a pseudo bulge in \so.  The 2D Fourier analysis and brightness asymmetry of the  spiral-arm like structure identified throughout our multi-filter imagery hints to a gravitational  interaction  origin.  Conversely, \citet{fisher_2008} did not find  conclusive evidence for a link between secular evolution and dynamical interaction.

If  we assume the bulge magnitude as  the most reliable proxy for the black hole mass,  then \so\ would represent a gamma-ray emitter which  black hole mass and luminosity are average among extragalactic high-energy sources. Prominent relativistic jets -- powerful enough to accelerate particles up to the highest energies--  have been always found to be launched from old red giant elliptical galaxies, thus linking the presence of relativistic jets to the latest stages of evolution.

However, the peculiar structure in the host galaxy of \so, in this work identified as a ring,  hints to  a recent violent dynamical interaction and might be taken as a  suggestive evidence of  a galaxy encounter as  responsible for the triggering of the AGN activity and  the initiation of a relativistic jet \citep[see][]{ramosalmeida_2012} responsible of the production of high-energy photons   Interestingly, while previous works have identified prominent radio jets being launched from  disk-like galaxies \citep[e.g.][]{vanbreugel_1984,heckman_1982,emonts_2008},  several of those galaxies show  a disturbed morphology likely associated to a  violent recent interaction, which comes in line with the interpretation of the \so\ host galaxy structure as a ring.

 This scenario is supported by the timescales  of galaxy interaction and radio structures derived  in \citet{anton_2008}. Therefore, it is likely  that  the gamma-ray nuclei of \so\ is hosted by a  galaxy that may have experienced an interaction  in the recent past, as suggested by the presence of the ring structure identified in this work.  If this picture holds, it would imply that a recently triggered AGN,  might  be able to launch (and collimate) fully developed relativistic outflows at an early evolutionary stage. This in turn would become an essential input to theoretical models for the formation of relativistic jets, AGN feedback and galaxy evolution \citep[e.g.][]{silk_1998,dimatteo_2005, hopkins_2005, debuhr_2010}.

We summarize this presentation by emphasizing two observational circumstances  regarding the host galaxy and nuclear activity of the gamma-ray  \nls\ \so: ($i$) The present multi-filter images do not allow us  to judge on the presence or absence of a  disc component in the host galaxy of \so. However,  the good representation of the surface brightness  of the \so\ host galaxy  by   \emph{model~A},  combination of  a bulge  and an unresolved  nuclear component,   presently does not argue for its presence. Nevertheless,  deeper and high resolution observations are  clearly needed to confirm the presence (or absence) of a disc component in \so.      ($ii$)  The discrepancy  between black hole mass estimates (via BLR radius and bulge magnitude)  can either suggest the presence of a flat distribution of the BLR or perhaps  a bulge that is still under a growing phase. The pole-on view of \so\ is supported by the one-sided jet structure observed and the detection of high energy photons. In the other hand,  we note that the resultant structural properties  of the bulge obtained in \emph{model A}  (e.g. $n \sim  2.8$) depart from the general properties of bulges launching prominent relativistic jets -- best fitted by a de Vaucouleurs profile (i.e.~$n = 4$).  Moreover, the extended structure in the host galaxy of \so, identified in this work as a  ring,  suggest a recent merger activity.    Unlike the general population of extragalactic gamma-ray sources (i.e. blazars), the host-galaxy of  \so\  does not seem to be passively evolving. The AGN activity in \so\ could  have been triggered by  inflows of gas provided by a recent galaxy interaction. Therefore, \so\ might reveal  itself as an outstanding case  where an AGN with a growing black hole might  be able to launch (and collimate) fully developed relativistic jets at an early evolutionary stage.

\section{Summary} 
 
Knowledge of the variability of the $\gamma$-ray \sy\  across the electromagnetic spectrum has increased substantially following recent detection of outburst episodes \citep[e.g.][]{abdo_2009,abdo_2009_monitoring,liu_2010,donato_atel_2011,foschini_2011,calderone_2011,dammando_2012,foschini_2012,jiang_2012,carpenter_2013, paliya_2013,dammando_1502_2013,dammando_0846_2013,Itoh_2013,tibolla_2013,wajima_2014}.  But the nature of the underlying galaxy that host the $\gamma$-ray \sy\ nuclei  has remained rather unexplored.  Since \sy\  are commonly hosted by spiral galaxies with high incidence of bars and ongoing star formation \citep{deo_2006, ohta_2007,sani_2010}, the detection of high-energy photons from \sy\  casts doubts on the exclusive relation between giant elliptical galaxies and  relativistic jets \citep[e.g.][]{kotilainen_1998_bllacs,kotilainen_1998_fsrq}.

In this context,  it is crucial to determine  what type of host galaxies  tend to harbor  $\gamma$-ray \sy. Hence, we have used  multi filter imagery and contemporaneous optical spectroscopy    to get a physical insight into the  host galaxy,  and  to estimate the black hole mass,   of the closest $\gamma$-ray \sy\ (\so~at~$z=0.061$) so far detected by  the \emph{Fermi Gamma-ray Space Observatory}. We summarize our results as follows:

\begin{enumerate}

\item Our NIR ($J$ and $K_s$) images of \so\  confirm  the irregular morphology  previously identified in optical images.  The $B, R, J, K_s$ surface brightness profiles of \so, have been decomposed into a sum of an AGN unresolved component~(PSF), S\'ersic~bulge and an exponential disk. We have used two models to perform the  surface brightness profile decomposition with GALFIT:  (\emph{model~A}) AGN+S\'ersic and  (\emph{model~B}) AGN+S\'ersic+Disk. The two models  are almost identical in terms  of $\chi^2$ reduced values (see Table~\ref{table:galfit})  and  smoothness of the   residual images. Since both models are equally adequate in describing the profile of \so\  within all four bandpasses, on statistical grounds,  we may adopt the simplest model (with less free parameters): \emph{model~A}.  
\item The  peculiar low surface brightness feature observed in the multi band images of   \so, has been  classified as a one-armed-spiral or a ring structure  by \citet{zhou_2007} and  \citet{anton_2008}, respectively. In order to shed some light on this apparent disagreement,  we have performed  a   2D Fourier analysis  aiming to characterize the dynamical structure  of the galaxy. The Fourier spectrum obtained from our analysis indicates that the feature seen in the images of \so, can be best represented by a ringed structure. We interpret this as  suggestive evidence for a recent violent dynamical interaction in \so,  likely related to the triggering of the AGN activity. If \emph{model~A} is correct, the ring has to be located inside the  bulge, presenting a rare and interesting case.

\item  The mass of the central black hole in \so,  estimated by   using the FWHM of the H$\beta$ emission lines measured from contemporary spectroscopy, is predicted to be about   $log~M_{BH}$~(R$_{BLR})\sim~7.2$. As a complementary approach, avoiding possible non-virial motions \citep{leontavares_2013}  and geometrical effects in the BLR \citep{decarli_2008},    we use the scaling relations between the mass of the black hole and the luminosity of the spheroid;   $log~M_{BH}$~(M$_{bulge})\sim8.6$ and $log~M_{BH}$~(M$_{bulge})\sim8.2$  for \textit{model~A} and \textit{model~B}, respectively. The black hole masses estimated by two  different scaling relations,   differ from each other  by  about one order of magnitude, which is more than  can be explained with the  measurement uncertainties and intrinsic scatter of the relations used. The black hole mass estimated via the bulge luminosity  is within a range of values typical of  $\gamma$-ray blazars, $\log M_{BH} = [7.8,9.2]$ \citep[e.g.][]{falomo_2002,falomo_2003,barth_2003,leontavares_2011_mbh}.

\end{enumerate}

Nevertheless, further integral-field unit (IFU) observations, as well as high spatial resolution, deep imaging observations  will  allow us to  confirm the host galaxy type of \so\ and  to properly authenticate the nature of such a peculiar structure as a ring -- remanent of a recent dynamical interaction. Based on the results presented in this work, it becomes evident that the attempt to characterize the host galaxy in other gamma-ray \sy\ is essential to understand the nature and evolutionary stage of this new class of gamma-ray emitters.  \\

We acknowledge the anonymous referee for  a very careful reading of the manuscript and comments that significantly improved this manuscript. This work was supported by CONACyT research grant 151494 (M\'exico). V. P. A. and A. O. I. acknowledge support from the CONACyT program for PhD studies. This paper is based on observations made with the Nordic Optical Telescope, operated on the island of La Palma jointly by Denmark, Finland, Iceland, Norway and Sweden, in the Spanish Observatorio del Roque de los Muchachos of the Instituto de Astrof\'{\i}sica de Canarias.

\bibliography{ms_05sep14}

\clearpage
\begin{turnpage}
\begin{deluxetable}{ccccccccccccccccccccccc}
\tablecolumns{16}
\tabletypesize{\scriptsize}
\tablecaption{Photometric and structural parameters of  \so\ host galaxy components\label{table:galfit}}
\tablehead{
\colhead{}  &\colhead{Psf} &\colhead{}  &  \multicolumn{4}{c}{Bulge} & \colhead{} &\multicolumn{3}{c}{Disk} &\colhead{} &\colhead{}\\
\cline{2-2} \cline{4-7} \cline{9-12} 
                    & \colhead{mag}&                   
                    & \colhead{mag} &  \colhead{$n$}  & \colhead{R$_{e}$} & \colhead{$<\mu_{e}>$}      &
                    &\colhead{mag} &  \colhead{R$_{s}$} & \colhead{axis ratio}&   
                    & \colhead{$\chi^{2}_{\nu}$}\\
                    & &                   
                    &  &    & [arcsec~/~kpc]  &   [mag arcsec$^{-2}$]&
                    & &  [arcsec/kpc] & & &   
                    & 
}
\startdata

\textbf{\emph{\footnotesize{Model~A} }}\\
\textbf{\emph{ \scriptsize{(PSF+Bulge)}} }\\
$B$        &15.71$\pm$ 0.01 &  &15.92$\pm$0.01&2.95$\pm$ 0.13&6.06$\pm$0.02~/~7.13$\pm$0.02& 21.57$\pm$0.01& & \nodata&\nodata& \nodata && 1.19$\pm$0.02 \\
$R$        &15.78$\pm$ 0.01 &  &15.13$\pm$0.01&2.65$\pm$0.07&4.66$\pm$0.03~/~5.48$\pm$0.03& 20.21$\pm$0.01& & \nodata&\nodata& \nodata  && 1.23$\pm$0.03\\
$J$         &14.25$\pm$ 0.01 &  &13.19$\pm$0.01&2.72$\pm$0.01&3.62$\pm$0.01~/~4.26$\pm$0.01& 17.72$\pm$0.01& & \nodata&\nodata& \nodata  && 1.31$\pm$0.02\\
$K_s$    &12.48$\pm$ 0.04 &  &12.00$\pm$0.01&2.87$\pm$0.01&2.70$\pm$0.01~/~3.18$\pm$0.01& 15.90$\pm$0.01 & & \nodata&\nodata& \nodata & & 1.29$\pm$0.01\\
\\

\textbf{\emph{\footnotesize{Model~B} }}\\
\textbf{\emph{ \tiny{(PSF+Bulge+Disk)}} }\\

$B$       &15.71$\pm$0.01&  &18.02$\pm$0.02&1.28$\pm$0.14&0.81$\pm$0.02/0.95$\pm$0.02&19.29$\pm$0.04&   & 16.38$\pm$0.02&3.22$\pm$0.02/3.79$\pm$0.02& 0.89$\pm$0.01&   & 1.20$\pm$0.01\\
$R$       &15.72$\pm$0.01&  &16.23$\pm$0.01&1.35$\pm$0.05&2.06$\pm$0.03/2.43$\pm$0.04&19.54$\pm$0.03&   & 15.80$\pm$0.02&4.53$\pm$0.04/5.33$\pm$0.05& 0.82$\pm$0.01&   & 1.22 $\pm$0.03\\
$J$        &14.18$\pm$0.01&  &14.65$\pm$0.01&0.88$\pm$0.01&1.25$\pm$0.01/1.48$\pm$0.01&16.89$\pm$0.01&   & 13.69$\pm$0.01&2.86$\pm$0.01/3.36$\pm$0.01& 0.94$\pm$0.01&   & 1.30$\pm$0.02\\
$K_s$   &12.44$\pm$0.01&  &13.03$\pm$0.01&1.24$\pm$0.01&1.19$\pm$0.01/1.40$\pm$0.01&15.14$\pm$0.01&   & 12.69$\pm$0.01&2.67$\pm$0.01/3.15$\pm$0.01& 0.96$\pm$0.01&   & 1.28$\pm$0.01\\

\enddata

\tablecomments{We adopt $H_{0} = 70$ Mpc$^{-1}$ km s$^{-1}$, $\Omega_{m}= 0.3$  and $\Omega_{\Lambda}=0.7$ cosmology,  which at  z=0.061 of \so\ yield  a scale of  1.177~kpc~arcsec${-1}$.   We compute errors for both models in an empirical way.  The best fit parameters are taken as initial guesses and we repeat the fitting procedure 10 times,  using different masks for each run. The masks vary in size and cover different objects and regions of the ringed arm structure.  Then, we compute the mean and standard deviation for every parameter and those are shown above. The error associated to the $\chi^{2}_{\nu}$ values is the standard deviation uncertainty  of the $\chi_{\nu}^{2}$ values obtained through the fitting procedure. }
\end{deluxetable}
\end{turnpage}
\clearpage

\clearpage
\begin{turnpage}
\begin{deluxetable}{l c c c c c c} 
\tablecolumns{6}
\tabletypesize{\scriptsize}
\tablecaption{Black hole mass estimates using H$\beta$ emission line \label{table:bh_spec}}
\tablehead{
\colhead{}  &\colhead{}              &\colhead{}             & \colhead{}               &\multicolumn{2}{c}{M$_{BH}(R_{BLR}$)}\\
\colhead{Model}                           &\colhead{FWHM(H$\beta$)} &\colhead{$ F_{\lambda}$(5100~{\rm\AA})}  & \colhead{{\it F}(H$\beta$)} & \colhead{$ L_{\lambda}$(5100~{\rm\AA})}   &\colhead{\it {L}(H$\beta$) }\\
\colhead{}                           &\colhead{(km s$^{-1}$)} &\colhead{(10$^{-17}$ erg s$^{-1}$ cm$^{-2}$ \AA$^{-1}$)}  & \colhead{(10$^{-15}$ erg s$^{-1}$ cm$^{-2}$)}  & \colhead{(log {\it M}$_{\odot}$)}   &\colhead{(log {\it M}$_{\odot}$)}\\
\colhead{(1)}                      &\colhead{(2)}                    & \colhead{(3)}  & \colhead{(4)}   &\colhead{(5)} & \colhead{(6)}
}
\startdata
\emph{a}  (1 Lorentzian) &  1301$\pm$160  &  113.1$\pm$8.1 & 161.8$\pm$10.1 & 6.9 &7.1\\
\emph{b}  (3 Gaussians) &  1660$\pm$190  &    91.0$\pm$5.8 & 163.3$\pm$10.3 & 7.1 &7.3\\
\enddata
\tablecomments{The reported FWHM are corrected by instrumental broadening, as calculated by measuring the width of an emission line near H$\beta$ in the HeAr spectra. The FWHM reported for the 3 Gaussian model, is the one measured from the sum of the two Gaussians that represent the broad component. The uncertainty of the black hole mass estimates is $\sim 0.4$ dex.}
\end{deluxetable} 
\end{turnpage}
\clearpage

\global\pdfpageattr\expandafter{\the\pdfpageattr/Rotate 90}

\end{document}